\documentclass[reprint, amsmath,amssymb,aps]{revtex4-2}

\usepackage{bbold}

\usepackage{graphicx}
\usepackage{dcolumn}
\usepackage{bm}
\usepackage[colorlinks=true, allcolors=blue,breaklinks=true]{hyperref}
\usepackage[dvipsnames]{xcolor}
\usepackage{nicefrac}
\usepackage{xr} 
\makeatletter
\newcommand*{\addFileDependency}[1]{
  \typeout{(#1)}
  \@addtofilelist{#1}
  \IfFileExists{#1}{}{\typeout{No file #1.}}
}
\makeatother

\newcommand*{\myexternaldocument}[1]{%
    \externaldocument{#1}%
    \addFileDependency{#1.tex}%
    \addFileDependency{#1.aux}%
}
\myexternaldocument{supplementary}

\DeclareMathOperator{\re}{Re}
\DeclareMathOperator{\imaginary}{Im}

\newcommand{\pd}{{\phantom\dagger}}
\providecommand{\abs}[1]{\lvert#1\rvert}

\begin{document}

\title{Detecting topological phases in the square-octagon lattice with statistical methods}

\author{Paul Wunderlich}
\email{wunderlich@itp.uni-frankfurt.de}
\author{ Francesco Ferrari}
\email{ferrari@itp.uni-frankfurt.de}
\author{Roser Valent\'{i}}
\email{valenti@itp.uni-frankfurt.de}
\affiliation{Institut f\"ur Theoretische Physik, Goethe Universit\"at Frankfurt am Main, Germany}

\date{\today}

\begin{abstract}
Electronic systems living on \textit{Archimedean} lattices such as kagome and square-octagon networks are presently being intensively discussed for the possible realization of topological insulating phases. Coining the most interesting electronic topological states in an unbiased way is however not straightforward due to the large parameter space of possible Hamiltonians. 
A possible approach to tackle this problem is provided by a recently developed statistical learning method [T. Mertz and R. Valent\'{i}, Phys. Rev. Research \textbf{3}, 013132 (2021)],  based on the analysis of a large data sets of randomized tight-binding Hamiltonians labeled with a topological index. In this work, we complement this technique by introducing a \textit{feature engineering} approach which helps identifying polynomial combinations of Hamiltonian parameters that are associated with non-trivial topological states. As a showcase, we employ this method to investigate the possible topological phases that can manifest on the square-octagon lattice, focusing on the case in which the Fermi level of the system lies at a high-order van Hove singularity, in analogy to recent studies of topological phases on the kagome lattice at the van Hove filling.
\end{abstract}

\maketitle

\section{Introduction}

The majority of structural and electronic properties of crystals are ultimately determined by the spatial arrangements of their atoms, which form a lattice structure that repeats periodically in space. In two dimensions, the enumeration of the lattice structures is connected to the problem of finding the possible \textit{tessellations} of a planar surface, i.e. the different ways to cover an infinite plane by a repeated juxtaposition of certain geometrical shapes (\textit{tiles}). When taking a single regular polygon as tile, only three possible networks which are homogeneous with respect to vertices, tiles and edges can be formed: the triangular, square and honeycomb lattices~\cite{chavey1989}. These periodic structures are usually referred to as \textit{Platonic} lattices and are ubiquitous in condensed matter systems. If the condition of homogeneity is loosened and one is allowed to employ different regular polygons as tiles, the so-called \textit{Archimedean} lattices can be constructed, which are homogeneous only with respect to vertices~\cite{footnote}.

\begin{figure}[t]
	\centering
	\includegraphics[width=\columnwidth]{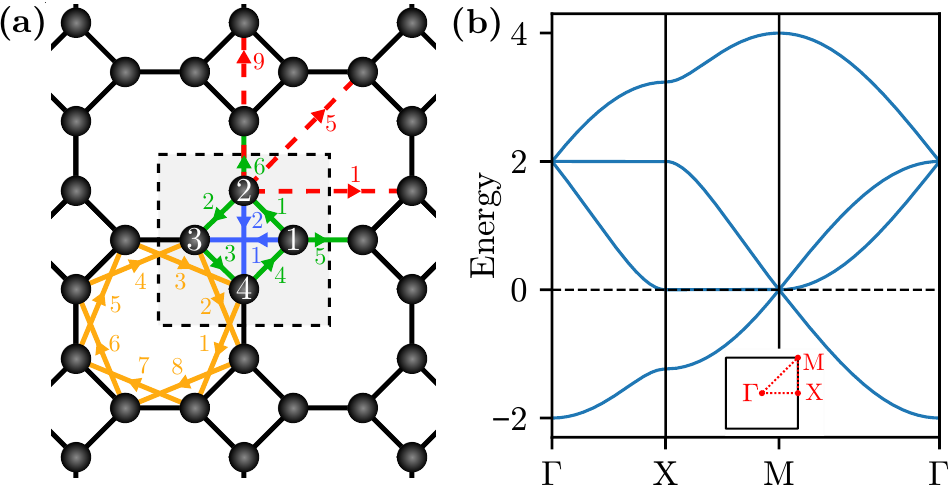}
	\caption[square-octagon lattice.]{(a)~Sketch of the square-octagon lattice. The unit cell is delimited by dashed black lines and contains four sites, labeled counterclockwise. Hopping terms are represented by colored lines: first-neighbor bonds in green ($t^1_{1\leq s \leq 6}$), second-neighbor bonds in blue ($t^2_{s=1,2}$), third-neighbor bonds in orange ($t^3_{1 \leq s \leq 8}$), fourth-neighbor bonds in red ($t^4_{1 \leq s \leq 12}$, dashed lines). For what concerns the latter, only three representative bonds are shown, for the sake of clarity. Arrows illustrate our convention for the complex-valued hopping terms, with $j \rightarrow i$ denoting the hopping parameter associated with the kinetic process $c_i^\dagger c_j$. 
	(b)~Band structure along the high symmetry path $\Gamma-$X$-$M$- \Gamma$ of the  first-neighbor tight-binding model for the square-octagon lattice, with uniform onsite terms $\epsilon^\pd_{1\leq s \leq4} = 1$ and uniform hopping parameters $t^n_{1\leq s \leq6} = -1$. The Fermi level has been set at $\nicefrac{1}{4}$ filling. The inset shows the Brillouin zone with its high symmetry points and lines in red.}
	\label{fig1} 
\end{figure}

The square-octagon lattice forms one of the eleven possible Archimedean tessellations of the two-dimensional plane~\cite{chavey1989}. It consists of a repetition of regular square and octagonal tiles, whose vertices define a crystal structure with a four-site unit cell repeated over an underlying square Bravais lattice [see Fig.~\ref{fig1}(a)]. The simple tight-binding treatment of the square-octagon nearest-neighbor network reveals rather intriguing properties of the electronic band structure: as shown in Fig.~\ref{fig1}(b), at $\nicefrac{1}{4}$ and $\nicefrac{3}{4}$ fillings, the energy dispersion shows a partially flat band intersecting two linearly dispersing bands, which form a Dirac cone. The flat dispersion results in the presence of a high-order van Hove singularity~\cite{Yuan2019}, namely a power-law divergence of the density of states, which is expected to enhance the effects of electronic correlations and favor the emergence of Fermi surface instabilities~\cite{PhysRevB.88.195104,PhysRevB.103.195104}. From a theoretical perspective, the question of the role of van Hove singularities for the onset of topological phases has been intensively investigated in the recent past in the context of the charge-density wave phase of AV$_3$Sb$_5$ kagome metals (with A=K, Rb, Cs)~\cite{kiesel2012,ortiz2019,neupert2022}. Several works suggested the existence of a topological flux phase among the possible instabilities of the kagome band structure at the van Hove filling~\cite{denner2021,park2021,lin2021,feng2021,2x2kagome}. It is worth noting that, although a simple tight-binding approach on the kagome lattice yields a band structure with conventional van Hove singularities, recent photoemission experiments detected the presence of a high-order van Hove singularity (close to the Fermi energy) in the band structure of CsV$_3$Sb$_5$~\cite{hu2022}. In this regard, the peculiar electronic dispersion of the square-octagon lattice is an intriguing minimal playground to investigate the possible onset of topological phases when the Fermi level of the system cuts through a high-order van Hove singularity.

It is worth mentioning that there are various proposals of two-dimensional compounds with a square-octagon geometry, such as monolayers of nitrogen group elements~\cite{zhang2015two}, metal nitrides and carbides~\cite{gaikwad2020octagonal},  a possible allotrope of monolayer MoS$_2$~\cite{PhysRevB.89.205402}, or two-dimensional polymers~\cite{springer2020,liu2021semimetallic}. Most importantly, several synthesis routes to fabricate T-graphene (octagraphene), a tetrasymmetrical carbon allotrope with a square-octagon periodic structure, have been put forward recently~\cite{enyashin2011,PhysRevLett.108.225505,sheng2012octagraphene,podlivaev2013kinetic,gu2019superconducting}. 
The square-octagon lattice is also found as a two-dimensional section of three-dimensional crystals, e.g. in the $xz$-plane of the \textit{Hollandite} structure, which is characteristic of certain Mn-oxides~\cite{luo2009,crespo2013a,crespo2013b,mandal2014,liu2014,maity2018}. Additionally, the one-fifth depleted square
lattice which describes the periodic arrangement of vanadium atoms in the antiferromagnetic CaV$_{4}$O$_{9}$ compound~\cite{taniguchi1995,katoh1995,kodama1997} is topologically equivalent to the square-octagon lattice (at first-neighbors) and often referred to as the CaVO lattice. In the past decades, several studies investigated Heisenberg-like models on the CaVO/square-octagon lattice in the context of frustrated magnetism~\cite{albrecth1996,troyer1996,ueda1996,starykh1996,sachdev1996,weihong1997,bao2014quantum,owerre2018,deb2020,maity2020}. On the other hand, more recently, a number of theoretical works have focused on different electronic models on the square-octagon lattice, with a focus on topological properties~\cite{PhysRevB.82.085106,PhysRevB.98.245116,liu2013topological,sil2019emergence,yang2019topological,liu2021semimetallic} and superconductivity~\cite{PhysRevB.99.184506,PhysRevB.101.224514}, mostly motivated by the synthesis of T-graphene~\cite{gu2019superconducting}.

In this work, we explore the possible topological phases that can manifest on the square-octagon lattice at $\nicefrac{1}{4}$ filling, i.e. when the Fermi level of the system lies precisely at the high-order van Hove singularity. Our study is based on a recently developed statistical learning method~\cite{Mertz_stat_method,mertz_thesis,2x2kagome}, in which a large data set of randomized tight-binding Hamiltonians is generated and subsequently analyzed by statistical tools drawn from machine learning approaches, with the purpose to gain insightful information on possible topological phases. We complement the methodology outlined in Ref.~\cite{Mertz_stat_method} by introducing \textit{feature engineering} as a tool to identify physical observables that are associated with non-trivial topological phases~\cite{mertz_thesis}. 

The paper is organized as follows:  Section~\ref{sec:method} is devoted to the description of the statistical method and discusses the concepts of marginal probability distributions, importance score, and feature engineering, which are employed for the data analysis; in Section~\ref{sec:model} we actualize the statistical method to the specific case of a square-octagon electronic system, defining the general form of the Hamiltonian; in Section~\ref{sec:results} we discuss the results of the statistical study, iterating several processes of dimensional reductions of the feature space in order to reach a minimal description of the topological phases; finally, in Section~\ref{sec:conclusions} we summarize our findings.

\section{Method}\label{sec:method}

Within the framework of the statistical method introduced in Ref.~\citep{Mertz_stat_method}, one can explore topological phases on an arbitrary lattice by considering fermionic tight-binding Hamiltonians of the following type
\begin{align}
	H = &\sum_i \epsilon_i^\pd c_i^\dagger c_i^\pd + \sum_{i,j} t_{i,j} c_i^\dagger c_j^\pd .
	\label{eq:tb_hamiltonian}
\end{align}
The two parts of the Hamiltonian consist of onsite potentials ($\epsilon_i \in \mathbb{R}$) and complex-valued hopping terms ($t_{i,j} \in \mathbb{C}$), respectively. The $t_{i,j}$ hopping integrals are taken to be translationally invariant and assumed to vanish when the distance between sites $i$ and $j$ exceeds a certain (arbitrary) threshold. For this reason, we conveniently introduce the notation $t^n_s$ for the various independent hopping terms, where $n$ runs over all possible Euclidean distances $R_n$, sorted in ascending order, and $s$ is an index denoting the inequivalent bonds at distance $n$. The electronic filling of the system is fixed by successively filling a certain number of energy bands. 

The independent onsite potentials and hopping parameters of the Hamiltonian are dubbed \textit{features} and collected into the vector $\vec{x}=(x_1,\dots,x_{N_f})$, with $N_f$ indicating the number of features. Within the statistical method, a certain choice of the entries of $\vec{x}$ is referred to as \textit{sample} and fully determines the electronic properties of the Hamiltonian. 
At a given filling, each sample can be characterized by the band gap $E_g$, which classifies samples into metals and insulators. 
If a sample is insulating, we can assign it a topological index, named \textit{label}, which distinguishes trivial and topological samples.  In the following, we choose the first Chern number $C$ as label~\citep{Berry,Zee_non_abel_C,Fukui_chern_algo}.

For an unbiased statistical analysis of a particular lattice, we generate a large number of different samples, i.e. different tight-binding Hamiltonian. 
This can be done by randomly picking tight-binding parameters $x_i$ according to a certain probability distribution function (PDF), e.g. a uniform or a Gaussian distribution. The generated data set is then analyzed by calculating the \textit{marginal probability distribution functions}
\begin{align}
    p_C(x_i) = \int \dots \int \rho_C(\vec{x}) \prod_{j\neq i} \text{d} x_j
\end{align}
for each feature $x_i$ and label $C$. Here, $\rho_C(\vec{x})$ is the bare PDF of all the \emph{insulating} samples with Chern number $C$. Therewith, by inspecting the properties of the marginal PDFs $p_C(x_i)$ of the various features $x_i$, we can determine the feature values which are most descriptive for the phase with Chern number $C$. This allows us to identify, for example, which patterns of hopping parameters is associated to a certain topological phase.

We note that features are in general complex-valued. For gaining most insight, one can examine the marginal PDFs for the real part ($\re[x_i]$), imaginary part ($\imaginary[x_i]$), modulus $(\abs{x_i})$ and phase ($\varphi[x_i] \equiv \arg[x_i]$) of each feature $x_i$.
The contrast between marginal PDFs for topological phases,  i.e. $p_{C\neq0}$, and marginal PDFs for trivial phases, i.e. $p_{0}$, indicates by which features a particular topological phase is characterized. In this regard, a quantitative measure of the importance of a certain feature $x_i$ for the topological phase with index $C$ is given by the Bhattacharyya distance~\citep{Bhattacharyya_distance} between the topological and trivial PDFs, namely
\begin{align}
    \label{eq:importance}
    D_B(p_{C\neq0}, p_{0}) = - \log\left[ \int_{\mathbb{C}} \sqrt{p_{C\neq0}(x_i) p_{0}(x_i)} \ \text{d} x_i \right].
\end{align}
This quantity, referred to as \textit{importance score} in the following, allows to perform a dimensional reduction of the feature space, by omitting the features with lowest values of $D_B(p_{C\neq0}, p_{0})$ in the course of the statistical analysis. It is worth mentioning that although other statistical distances between probability distributions can be adopted for the definition of the importance score (e.g., the Hellinger distance~\cite{pardo2005}), a previous benchmark study on the honeycomb lattice has shown that the Bhattacharyya distance provides a better contrast of the marginal distributions with respect to other metrics~\cite{mertz_thesis,Mertz_stat_method}.
Complementary to the use of the importance score, a model can be refined by establishing symmetries between features, as obtained either from physical grounds or from the behavior of the marginal PDFs~\cite{2x2kagome}. An iterative application of the statistical method, involving subsequent data generation, dimensionality reduction and analysis of the marginal PDFs, leads to the definition of effective models for topological phases. 

Furthermore, in the present work we pursue a better understanding of the parametrization of topological phases by introducing a \textit{feature engineering} procedure. We define additional composite features by taking certain combinations, e.g. sums, products or power series, of (some of) the original features and compute their corresponding importance score as the Bhattacharyya distance between the trivial and topological marginal PDFs. Some of these engineered features may carry higher importance score and serve as particularly outstanding descriptors of a particular phase. 

Employing the statistical method outlined in this section, complemented by feature engineering, we tackle the study of topological states that can manifest in the square-octagon lattice.

\section{Lattice and model} \label{sec:model}

The square-octagon lattice, sketched in Fig.~\ref{fig1}~(a), is defined by a square Bravais lattice and a unit cell of four lattice sites. Denoting the Bravais lattice vectors by $\mathbf{a}_1= (1,0)$ and $\mathbf{a}_2= (0,1)$, the four sites inside the unit cell can be placed at positions $\pm \nicefrac{\sqrt{2}}{2}\ \mathbf{a}_1$ and $\pm \nicefrac{\sqrt{2}}{2} \ \mathbf{a}_2$.
To investigate possible topological phases on this lattice we consider a spinless tight-binding Hamiltonian of the form of Eq.~\eqref{eq:tb_hamiltonian}, with hopping terms being restricted from first to fourth-neighboring sites. Assuming translational invariance, the model contains four onsite potentials, with parameters $\epsilon_{1\leq s\leq 4}$, and a total of $28$ hopping parameters. 
As shown by the different colored lines in Fig.~\ref{fig1}~(a), the $28$ hoppings are divided into six first-neighbor terms $t^1_{1\leq s\leq6}$ (green lines), two second-neighbor terms $t^2_{s=1,2}$ (blue lines), eight third-neighbor terms $t^3_{1\leq s \leq 8}$ (orange lines), and twelve fourth-neighbor terms $t^4_{1\leq s\leq 12}$ (dashed red lines; for the sake of clarity, only three symmetry-inequivalent links are shown).

The band structure for the model with uniform onsite terms, ${\epsilon_{1\leq s\leq 4} = 1}$, and uniform first-neighbors hoppings, ${t^1_{1\leq s\leq6} = -1}$ (${t^{n>1}_s=0}$), is shown in Fig.~\ref{fig1}~(b). The system is metallic for any filling. The dashed horizontal line indicates the Fermi energy at $\nicefrac{1}{4}$ filling, where the dispersion is characterized by a triply degenerate point at $M$, formed by the lowest-lying three bands and consisting of a Dirac node and a (partially) flat band. Another band crossing of the same type, formed by the upper three bands, occurs at the $\Gamma$ point.

By tuning the hopping parameters it is possible to create topological insulators, i.e., open a gap in the energy bands and induce non-zero Chern numbers. In the following, we will infer which parameters have to be manipulated in order to create topological insulators.
We focus on the case of $\nicefrac{1}{4}$ filling, for which the Fermi energy intersects the lower triply degenerate band crossing. The presence of partially flat bands gives rise to high-order van Hove singularities in the density of states~\cite{PhysRevB.103.195104}, which implies a strong susceptibility of the system towards symmetry breaking in the presence of electron-electron interactions~\cite{Yuan2019,PhysRevB.88.195104}. The Fermi surface of the system coincides with the edges of the Brillouin zone, i.e. it can be seen as the square connecting the $M$ points. Along its vertical (horizontal) edge, i.e. $\mathbf{k}=(\pi,k)$ [$\mathbf{k}=(k, \pi)$], the Fermi surface displays a mixed sublattice character, with the Bloch waves being evenly localized on sublattice sites $1$ and $3$ ($2$ and $4$).

\section{Statistical analysis}~\label{sec:results}

\begin{figure}[t]
	\centering
	\includegraphics[width=\columnwidth]{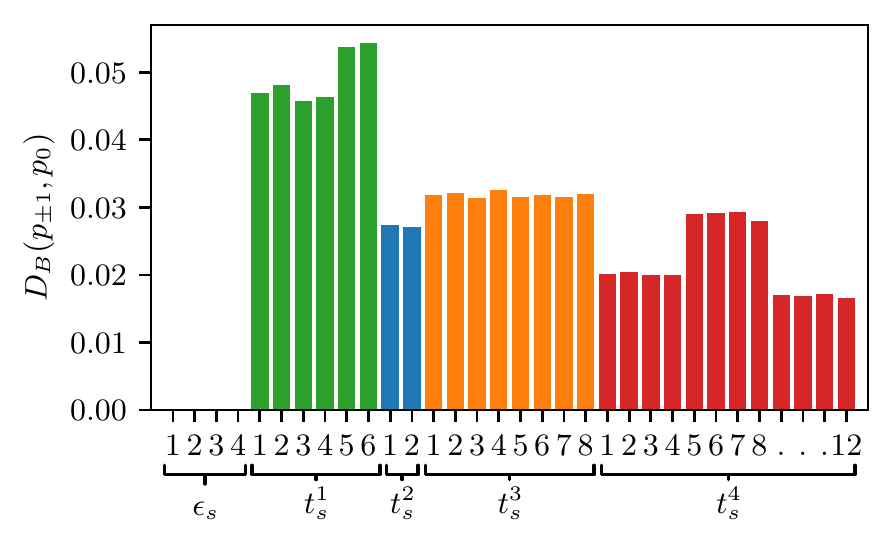}
	\caption{Importance score $D_B(p_{\pm 1}, p_0)$ [(see Eq.~\eqref{eq:importance}] for all features of the square-octagon tight-binding model, namely onsite terms and hoppings up to fourth neighbors.}
	\label{fig2}
\end{figure}

For the statistical analysis of the square-octagon lattice we begin by considering the tight-binding model of Eq.~\eqref{eq:tb_hamiltonian} with hopping terms up to fourth-neighbor bonds. In order to randomly sample the feature space, we define a set of reference values for each feature, generally denoted as $x_i^\text{ref}$~\cite{Mertz_stat_method}. The samples are drawn according to a multivariate (two-dimensional) Gaussian distribution in the complex plane, centered in $x_i^\text{ref}\in \mathbb{C}$ and with covariance matrix ${\Sigma=\alpha^2\abs{x_i^\text{ref}}^2 \mathbb{1}_{2\times2}}$, where $\alpha \in \mathbb{R}$ is an arbitrary hyperparameter. Analogously, for real-valued features, i.e. onsite potentials, a one-dimensional Gaussian is employed. As reference points for the various features, we take ${\epsilon_{1 \leq s \leq 4}^\text{ref} = 0.25}$, ${t^{1, \text{ref}}_{1\leq s\leq6} = -1}$, ${t^{2, \text{ref}}_{s=1,2} = \nicefrac{-1}{\sqrt{2}}}$, ${t^{3, \text{ref}}_{1\leq s\leq8} = \nicefrac{-1}{\sqrt{2+\sqrt{2}}}}$ and ${t^{4, \text{ref}}_{1\leq s\leq8} = \nicefrac{-1}{(1+\sqrt{2})}}$. Note that the reference points of the hopping terms are scaled by the inverse distance between $n$th neighbors, i.e. $1/R_n$. For the width of the Gaussian PDFs, we take $\alpha=0.6$. This scheme allows us to consider physical Hamiltonians where extreme values of tight-binding parameters are excluded~\cite{Mertz_stat_method}. For example, within our parametrization, the choice $\alpha=0.6$ ensures that the real part of the extracted features does not change sign with respect to the reference point for most samples ($\approx 95\%$). We verified that small changes of $\alpha$ with respect to the above choice do not affect the results significantly. However, in general, extreme values of $\alpha$ shall be avoided. Indeed, if $\alpha$ is too small the sampling is limited to Hamiltonians which are close to the reference point and does not cover a significant amount of the feature space; on the other hand, for a fixed number of samples, choosing a larger value of $\alpha$ leads to noisier marginal PDFs, which may hamper the statistical analysis. We note that the choice of the reference point constitutes the main bias of the present approach. The simplest way to alleviate this bias involves choosing different initial reference points to cover a larger portion of the feature space. The choice can be based on an iterative application of the statistical method: once a set of parameters yielding a certain topological phase is identified, one can perform a new statistical analysis centered around the topological reference point, thus exploring the feature space around it. On the other hand, biasing the results around a certain reference point can be desirable in the case in which the present method is applied to a specific physical system. For example, if one is interested in exploring topological phases for a certain target material, the reference point can be chosen to be an \textit{ab-initio} determined tight-binding Hamiltonian~\cite{mertz_thesis}. 

After creating a data set of $n_S = 2 \cdot 10^7$ samples on the square-octagon lattice, we find 3.3\% insulators out of all samples, 17.6\% of which are topological. 
Nearly all topological insulators (99.6\%) have Chern index $C=\pm 1$. As we sample in a large parameter space, the number of topologically non-trivial samples is small. 
Hence, we proceed attempting a dimensional reduction in order to infer more information on the topological phases.

\subsection{Dimensional reduction}
\label{sec:dim_reduction}

\begin{figure*}[t]
\centering
\includegraphics[width=0.9\textwidth]{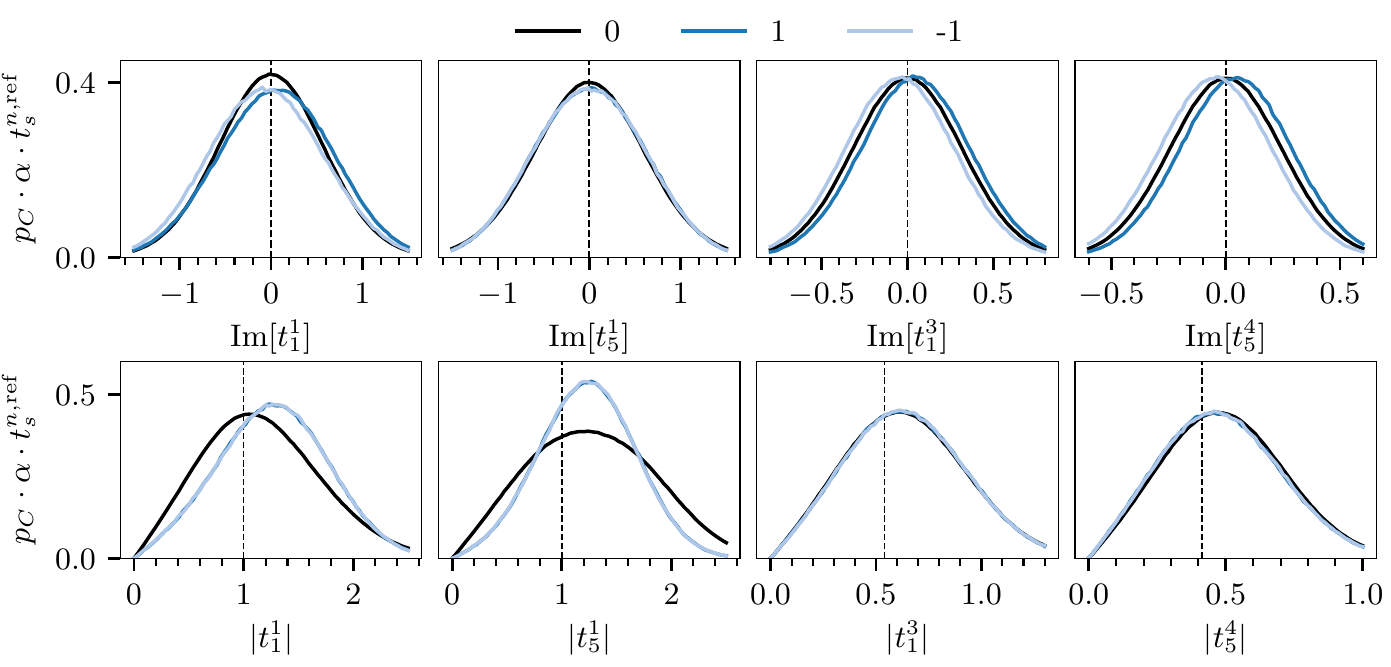}
\caption{Marginal PDFs for the imaginary part $\imaginary[t^n_s]$ and modulus $\abs{t^n_s}$ of representative features of the tight-binding model on the square-octagon lattice with first-, third- and fourth neighbor hoppings (the latter being restricted to those connecting sites of different sublattices). A sketch of the bonds corresponding to the different panels is given in Fig.~\ref{fig1}~(a). PDFs for the trivial phase ($C=0$) are indicated by gray lines, while PDFs for topological phases ($C\neq0$) are colored in shades of blue, as indicated in legend at the top of the figure. The reference value of each feature is marked by a dashed vertical line. The PDFs have been rescaled by a factor $\alpha t_s^{n,\text{ref}}$ only for visualization purposes.}
\label{fig3}
\end{figure*}

The parameter space can be reduced by examining the importance scores $D_B(p_{1}(x_i),p_0(x_i))$ and $D_B(p_{-1}(x_i),p_0(x_i))$ for the $C=\pm1$ phases, which constitute the majority of topological samples.
The importance score of each feature is the same for $C=1$ and $C=-1$, because the underlying marginal PDFs show specular behavior with respect to the $\imaginary(x_i)=0$ axis in complex plane for opposite Chern numbers. 
As shown in Fig.~\ref{fig2}, we observe zero importance for onsite terms.
Hence, the parameters $\epsilon_i$ do not play any role in distinguishing trivial and topological phases, and can thus be excluded from the statistical analysis.
On the other hand, the importance score of all hopping parameters is finite. 

Similar values of the importance scores, which vary up to statistical noise due to the finite sample count of topological insulators, indicate the presence of sub-groups of hoppings, as expected from the inherent symmetry of the square-octagon lattice. Indeed, we can distinguish two classes of first-neighbor hoppings, according to their importance score: (i)~bonds within square plaquettes $\{t^1_1, t^1_2, t^1_3, t^1_4\}$ and (ii)~bonds connecting square plaquettes $\{t^1_5, t^1_6\}$. Also fourth neighbor hoppings can be grouped in three classes: (i)~bonds crossing the square plaquettes $\{t^4_1, t^4_2, t^4_3, t^4_4\}$ [vertical red dashed line in Fig.~\ref{fig1}~(a)], (ii)~bonds crossing the octagonal plaquettes and connecting sites belonging to the same sublattice $\{t^4_5, t^4_6, t^4_7, t^4_8\}$ [horizontal red dashed line in Fig.~\ref{fig1}~(a)] and (iii)~bonds crossing the octagonal plaquettes and connecting sites belonging to different sublattices $\{t^4_9, t^4_{10}, t^4_{11}, t^4_{12}\}$ [diagonal red dashed line in Fig.~\ref{fig1}~(a)]. For what concerns second-neighbor hoppings ($t^2_{s=1,2}$) and third-neighbor hoppings ($t^3_{1\leq s \leq 8}$) no distinction into sub-groups can be made based on the importance score.

Among all hoppings, the lowest importance is shown by the fourth-neighbor hoppings which connect sites belonging to the same sublattice, i.e. classes (i) and (ii). Therefore, based on this observation, we omit these parameters (and the onsite potentials) in the next iteration of our analysis. A new sampling procedure with the reduced model yields a remarkably larger number of insulators (64\% of all samples) and shows a rather low importance for the second-neighbor hopping terms, which is approximately three times lower than the importance of third- and fourth-neighbor terms (not shown). Hence, based on this observation, we also exclude the second-neighbor hoppings from our model, in order to scale down the size of the feature space. This will enhance the contrast between the marginals PDFs and thus simplify the subsequent analysis.

We are thus left with a model including only first-, third-, and fourth-neighbor hoppings of class (iii) (i.e., those connecting sites belonging to different sublattices). Note that, for simplicity, we will refer to the latter as "fourth-neighbor hoppings" in the remainder of the paper.
The new data set contains 64\% insulators, 18.2\% of which possess a non-trivial Chern index $C=1$ or $C=-1$.
The fraction of insulators with higher Chern number is negligibly small. 
Compared to the previous iterations, we observe a higher portion of topological insulators due to the reduced parameter space (11.6\% out of all samples, against 0.57\% for the full model including onsite terms and all hoppings up to fourth-neighbors). 

We can gather information on topological phases from this model by considering the marginal probability distributions.
Based on their appearance, the PDFs of first-neighbor hoppings can be grouped in two subsets, one formed by the hoppings inside the square plaquettes \{$t^1_{1}$, $t^1_{2}$, $t^1_{3}$, $t^1_{4}$\}, and the other containing hoppings that connect adjacent plaquettes \{$t^1_{5}$, $t^1_{6}$\}, see Fig.~\ref{fig1}~(a).
The marginal distributions for third- and fourth-neighbors, respectively, show the same behavior among each type.
One exemplary set of the PDFs of imaginary parts $p_C(\imaginary[t^n_s])$, which indicate the ``directions'' of the complex hoppings, and PDFs of the moduli $p_C(\abs{t^n_s})$, which describe the overall hopping strengths, is shown in Fig.~\ref{fig3} for each group of hoppings.
From these PDFs we can infer the most descriptive features characterizing trivial and topological phases, as discussed in the following.

\subsubsection{Trivial $C=0$ phase}

In the trivial phase, the marginal PDFs for the imaginary parts of all hoppings shown in Fig.~\ref{fig3}, i.e., $p_0(\imaginary[t^1_1])$, $p_0(\imaginary[t^1_5])$, $p_0(\imaginary[t_1^3])$ and $p_0(\imaginary[t_5^4])$ show a perfect symmetric behavior around zero. We can thus infer that no specific hopping direction is preferred. The
PDFs of the moduli $p_0(\abs{t^1_1})$, $p_0(\abs{t^1_5})$, $p_0(\abs{t_1^3})$ and $p_0(\abs{t_5^4})$ show similar shapes, with a non-zero mean indicating finite bond strengths. Hence, the trivial insulating phase can be realized by finite first-, third- and/or fourth-neighbor hoppings, with no specific hopping direction (e.g., by real hoppings). This configuration is schematically illustrated in the left panel of Fig.~\ref{fig4}, where we color the relevant bonds within one unit cell.

\subsubsection{Topological $C=\pm1$ phases}

For first-neighbor hoppings within the square plaquettes, exemplified by the term $t^1_{1}$ in Fig.~\ref{fig3}, we observe that the marginal PDFs for non-zero Chern numbers, i.e., $p_{\pm1}(\imaginary[t^1_1])$ and $p_{\pm1}(\abs{t^1_1})$, look rather distinct from the PDFs of the trivial phase.
For the $C=1$ phase, $\imaginary[t^1_1]$ tends to be larger than zero which corresponds to a counter-clockwise winding of the hoppings around the square plaquettes. 
$p_{-1}(\imaginary[t^1_1])$ is the conjugate of $p_{1}(\imaginary[t^1_1])$, hence the winding is clockwise.
The modulus $\abs{t_1^1}$, i.e., the overall hopping strength, shows larger values for topological phases than for the trivial phase.
For what concerns the remaining first-neighbor hoppings, represented by the term $t^1_{5}$ in Fig.~\ref{fig3}, we observe that $p_{\pm1}(\imaginary[t^1_5])$ is symmetric around zero, i.e., no particular hopping direction is indicated and, thus, these hoppings do not play a role in differentiating between $C =\pm1$ and $C=0$ phases.
At variance with the case of $\abs{t_1^1}$, the marginal PDFs of $\abs{t_5^1}$ have similar means for $C=\pm1$ and $C=0$ phases.

As shown in Fig.~\ref{fig3} by the representative term $t^3_{1}$, also the marginal PDFs for the imaginary part of third-neighbor hoppings behave differently for topological and trivial phases:
 $\imaginary[t^3_1]$ tends to be larger than zero for $C=1$, while for $C=-1$ it shows a tendency to be smaller than zero.
This implies that non-zero third-neighbor bonds with complex hoppings winding clockwise (anti-clockwise) in the octagonal plaquettes can support the non-trivial $C=1$ ($C=-1$) phase.
On the contrary, the marginal PDFs of the moduli $p_C(\abs{t^3_1})$ look identical for $C=\pm1$ and $C=0$ and, thus, they provide no information about the topological properties. Finally, we observe that the marginals of fourth-neighbor hoppings, represented by $t_5^4$, show qualitatively the same behavior as the marginals for the third-neighbor bonds. 
Hence, also non-zero fourth-neighbor bonds with hopping directions winding clockwise (anti-clockwise) can support the non-trivial $C=1$ ($C=-1$) phase.

\begin{figure}[t]
\centering
\includegraphics[width=\columnwidth]{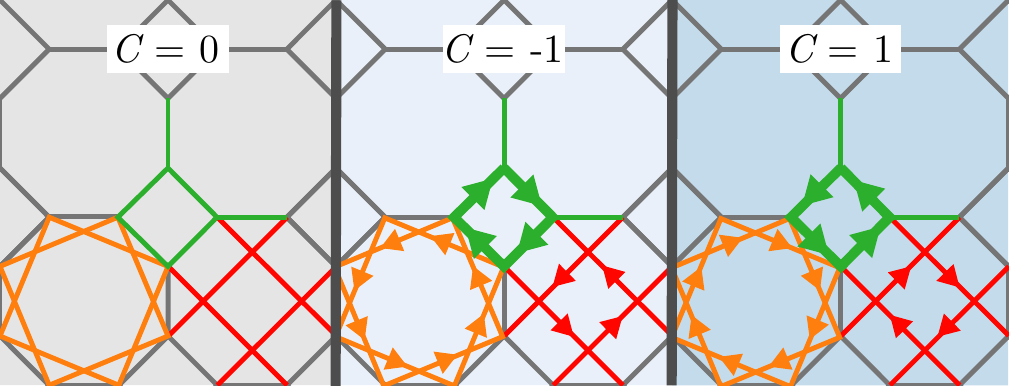}
\caption{Schematic illustration of possible trivial and topological phases of the tight-binding model on the square-octagon lattice. First-neighbor hoppings are colored in green, third-neighbor hoppings in orange and fourth-neighbor hoppings in red. The thickness of the lines indicates the relative bond strength given by $\abs{t^n_s}$. Arrows indicate the hopping direction as given by $\imaginary[t^n_s]$.}
\label{fig4}
\end{figure}

In summary, a topologically insulating phase with $C=1$ can be induced by anti-clockwise first neighbor hoppings on the square plaquettes, which are relatively stronger than the bonds connecting square plaquettes, together with clockwise third and fourth neighbor hoppings. 
Topological insulators with ${C=-1}$ can be created by reversing the hopping directions of the $C=1$ phase.
These results are schematically summarized by the sketches in the middle and right panel of Fig.~\ref{fig4}. Here, the thickness of the bonds reflects the relative hopping strengths and the arrows illustrate the hoppings directions, i.e., the sign of their imaginary parts.

\begin{figure}
\centering
\includegraphics[width=\columnwidth]{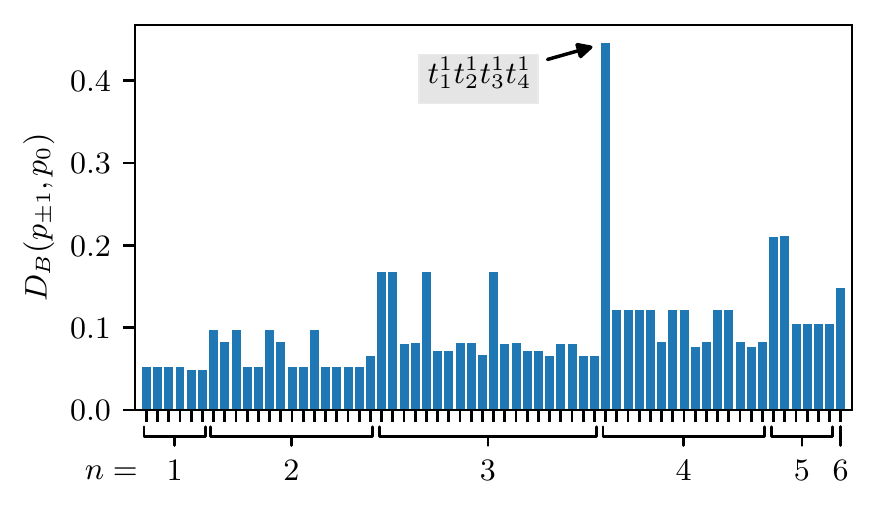}
\caption{Bhattacharrya distance $D_B(p_{\pm1}, p_0)$ as given by Eq.~\eqref{eq:importance} for the first neighbor hoppings $t^1_{1\leq s\leq6}$ and all engineered features which were constructed by taking products of $n$ distinct first-neighbor hoppings.}
\label{fig5}
\end{figure}

\subsection{Towards a first-neighbor model and feature engineering}

\begin{figure*}[t]
	\centering
    \includegraphics[width=1\textwidth]{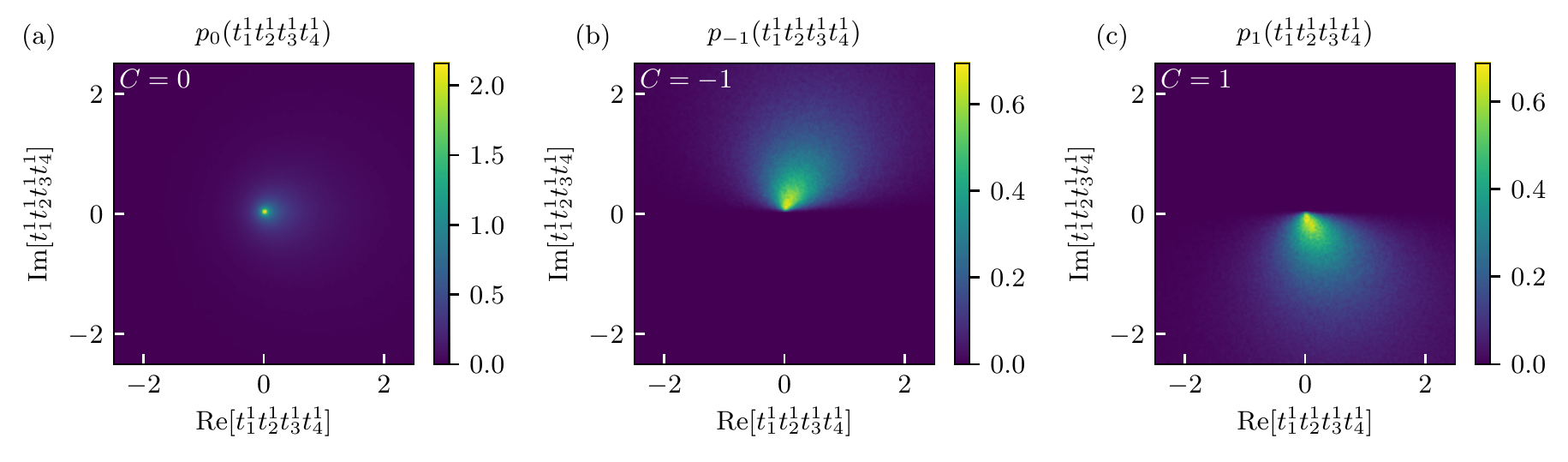}
    \caption{Marginal PDFs in complex space for the engineered feature $t^1_1 t^1_2 t^1_3 t^1_4$ of the first-neighbor tight-binding model on the square-octagon lattice for the (a)~$C=0$ trivial phase, (b)~$C=-1$ topological phase and (c)~ $C=1$ topological phase.}
	\label{fig6}
\end{figure*}

To gain a deeper understanding of the phases which can manifest in the square-octagon lattice, we continue with a reduction of parameters based on the importance scores.
Within the tight-binding model with first-, third- and fourth-neighbor hoppings discussed in the previous section, the importance score of first-neighbor terms turns out to be up to eight times larger than the importance score of third- and fourth-neighbor terms.
Based on these observations we exclude third and fourth neighbor terms as the next step of our analysis.

Creating a data set for the Hamiltonian with only first-neighbor hoppings yields 95.5\% insulating samples. The fraction of topological insulators corresponds to 13.7\% of all samples, analogously to what has been observed in the calculation with first-, third- and fourth-neighbor hoppings. This further indicates the higher importance of first-neighbor hoppings for the topologically non-trivial phases with ${C=\pm 1}$.
With the first-neighbor model we arrive at the minimal possible description for topological phases on the square-octagon lattice.
As done previously, we can group first-neighbor hoppings into subsets based on the behavior of marginal PDFs (not shown): (i) hoppings that form square plaquettes, $\{t^1_1, t^1_2, t^1_3, t^1_4\}$, and (ii) hoppings connecting different squares $\{t^1_5, t^1_6\}$. 

In order to try gaining additional information on the topological phases, we apply feature engineering, namely we define new composite features by taking all possible products involving (distinct) first-neighbor hoppings, i.e. pair-wise products of the form $t_s^1 t_{s'}^1$, triple products of the form $t_s^1 t_{s'}^1 t_{s''}^1$, and so on, up to the product of all six first-neighbor hoppings. We then calculate the importance score for the newly engineered features and identify the ones which play a major role in characterizing the topological phases.

As shown in Fig.~\ref{fig5}, the product of all hoppings on the square plaquettes, namely $t^1_1 t^1_2 t^1_3 t^1_4$, turns out to possess a remarkably large importance score, $D_B = 0.45$ (c.f. $D_B \approx 0.05$ for $ t^1_{1\leq s \leq 6}$). 
For this particular engineered feature, the marginal PDFs in the complex plane, shown in Fig.~\ref{fig6}, provide crucial insight.
In the $C=0$ phase, the PDF of $t_1 t_2 t_3 t_4$ is symmetric with respect to the real axis, as shown in Fig.~\ref{fig6}~(a). 
On the other hand, the marginals for the topological phases [Fig.~\ref{fig6}~(b) and (c)] are completely localized in the upper and lower half of the complex plane for $C=-1$ and $C=1$, respectively.
Hence, the importance score for distinguishing the two topological phases takes its maximal value, i.e., $D_B(p_1(t^1_1 t^1_2 t^1_3 t^1_4), p_{-1}(t^1_1 t^1_2 t^1_3 t^1_4)) = \infty$. 
This implies that the distinct topological phases are unambiguously distinguished by this engineered feature.
Physically, the topological phases are distinguished by the phase picked up after one loop in the square plaquette which is given by $\varphi[t^1_1 t^1_2 t^1_3 t^1_4] = \varphi[t^1_1] + \varphi[t^1_2] + \varphi[t^1_3] + \varphi[t^1_4]$. 
Eventually, the engineered feature $t^1_1 t^1_2 t^1_3 t^1_4$ may serve as the unique descriptor of the topological phases.

\section{Summary}\label{sec:conclusions}

The statistical method introduced in Ref.~\cite{Mertz_stat_method} constitutes an effective procedure to identify possible topological phases that can be realized by a tight-binding Hamiltonian on a given lattice structure. We employed this technique to scrutizine the topological phases appearing at the high-order van Hove filling on the square-octagon lattice, which forms one of the eleven \textit{Archimedean} tessellations of the two-dimensional Euclidean plane and is realized in a number of different materials. 
Starting from a generic tight-binding model with hoppings up to fourth nearest neighbors, we constructed a dataset of randomized Hamiltonians labelled by their Chern number as topological index. We then performed a statistical analysis of the marginal probability distributions for the various parameters of the system and, by means of dimensional reduction, we reached an effective model describing topological phases with Chern number $C=\pm 1$ on the square-octagon lattice. Most importantly, we introduced a \textit{feature engineering} procedure that allows us to gain deeper insight into the nature of the topological phases by identifying polynomial combinations of tight-binding parameters which are associated to non-trivial topology, e.g. Peierls-like fluxes. Going beyond the methodological improvements presented in this work and the results for the square-octagon lattice, the present statistical method can be regarded as a potential tool to perform a material-specific search of topological phases, by exploring the phase space around a tight-binding Hamiltonian obtained from first principles (e.g. by density-functional theory and Wannierization). The search of topological phases and its characterization by means of engineered features could serve as a guide to experimental manipulation of the target material to tune its properties towards desired topological phases, e.g. by means of applied pressure or strain.

\acknowledgments{We would like to thank Thomas Mertz for valuable inputs and discussions. We acknowledge support from the Deutsche Forschungsgemeinschaft (DFG, German Research Foundation) through TRR 288-422213477 (Project B05) and through QUAST FOR 5249-449872909 (Project P4).}


\begin{thebibliography}{57}
\bibitem [{foo()}]{footnote}%
  \BibitemOpen
  \href@noop {} {}\bibinfo {note} {In other words, each vertex of these
  networks is surrounded by the same set of regular polygons.}\BibitemShut
  {Stop}%
\end{thebibliography}
\end{document}